\documentclass[aps,prd,nofootinbib,amsmath,amssymb,superscriptaddress,showpacs,twocolumn,9pt]{revtex4}
\usepackage{txfonts}
\usepackage{graphicx}
\usepackage{dcolumn}
\usepackage{bm}
\usepackage{amssymb}
\usepackage{latexsym}
\usepackage{booktabs}
\usepackage[colorlinks=true, linkcolor=red, citecolor=blue]{hyperref}

\newcommand{\be}{\begin{equation}}
\newcommand{\ee}{\end{equation}}
\newcommand{\bq}{\begin{eqnarray}}
\newcommand{\eq}{\end{eqnarray}}

\begin{document}

\title{New initial condition of the new agegraphic dark energy model}

\author{Yun-He Li}
\affiliation{Department of Physics, College of Sciences,
Northeastern University, Shenyang 110004, China}
\author{Jing-Fei Zhang}
\affiliation{Department of Physics, College of Sciences,
Northeastern University, Shenyang 110004, China}
\author{Xin Zhang\footnote{Corresponding author}}
\email{zhangxin@mail.neu.edu.cn} \affiliation{Department of Physics,
College of Sciences, Northeastern University, Shenyang 110004,
China} \affiliation{Center for High Energy Physics, Peking
University, Beijing 100080, China}

\begin{abstract}
The initial condition $\Omega_{\rm de}(z_{\rm ini})=n^2(1+z_{\rm
ini})^{-2}/4$ at $z_{\rm ini}=2000$ widely used to solve the
differential equation of the density of the new agegraphic dark
energy (NADE) $\Omega_{\rm de}$, makes the NADE model be a
single-parameter dark-energy cosmological model. However, we find
that this initial condition is only applicable in a flat universe
with only dark energy and pressureless matter. In fact, in order to
obtain more information from current observational data, such as
cosmic microwave background (CMB) and baryon acoustic oscillations
(BAO), we need to consider the contribution of radiation. For this
situation, the initial condition mentioned above becomes invalid. To
overcome this shortage, we investigate the evolution of dark energy
in the matter-dominated and radiation-dominated epochs, and obtain a
new initial condition $\Omega_{\rm de}(z_{\rm
ini})=\frac{n^2(1+z_{\rm ini})^{-2}}{4}\left(1+\sqrt{F(z_{\rm
ini})}\right)^2$ at $z_{\rm ini}=2000$, where
$F(z)\equiv\frac{\Omega_{r0}(1+z)}{\Omega_{m0}+\Omega_{r0}(1+z)}$
with $\Omega_{r0}$ and $\Omega_{m0}$ being the current density
parameters of radiation and pressureless matter, respectively. This
revised initial condition is applicable for the differential
equation of $\Omega_{\rm de}$ obtained in the standard
Friedmann-Robertson-Walker (FRW) universe with dark energy,
pressureless matter, radiation, and even spatial curvature, and can
still keep the NADE model being a single-parameter model. With the
revised initial condition and the observational data of type Ia
supernova (SNIa), CMB and BAO, we finally constrain the NADE model.
The results show that the single free parameter $n$ of the NADE
model can be constrained tightly.
\end{abstract}

\pacs{95.36.+x, 98.80.Es, 98.80.-k} \maketitle


\section{Introduction}\label{sec1}

The accelerated expansion of current universe, first observed in
1998~\cite{Riess98}, implies that our universe is being dominated by
an exotic component with negative pressure dubbed dark energy. To
understand its nature, we should first ascertain its dynamic
evolution. For many dark energy models, it is believed that the
models are favored by observations if they can fit the data well
with less free parameters, since the less parameters may be
constrained tightly. To our knowledge, in dark-energy cosmology,
there exist three rare dark energy models, namely, the Lambda Cold
Dark Matter ($\Lambda$CDM)~\cite{lcdm}, the Dvali-Gabadadze-Porrati
(DGP) braneworld~\cite{Dvali:2000hr}, and the new agegraphic dark
energy (NADE)~\cite{Wei:2007ty} models, which contain only one free
model parameter. Among these three models, the NADE model is a
special one, because unlike the other two models whose
single-parameter feature is obvious, the NADE model is due to its
special analytic feature in the matter-dominated
epoch~\cite{Wei:2007xu}. To see it clearly, we first briefly review
the NADE model.

The dark energy density $\rho_{\rm de}$ in the NADE model,
constructed in light of the K\'{a}rolyh\'{a}zy
relation~\cite{uncertainty} and corresponding energy fluctuations of
space-time, has the form~\cite{Wei:2007ty}
 \be\label{eq:rhoq}
   \rho_{\rm de}=\frac{3n^2M_{\rm Pl}^2}{\eta^2},
 \ee
where $n$ is a numerical parameter, $M_{\rm Pl}$ is the reduced
Planck mass. The $\eta$ is the conformal age of the universe
 \be\label{eq:eta}
   \eta\equiv\int_0^t\frac{dt'}{a}=\int_0^a\frac{da'}{Ha'^2},
 \ee
where $a$ is the scale factor of the universe, and
$H\equiv\dot{a}/a$ is the Hubble parameter. Here the overdot denotes
the derivative with respect to the cosmic time $t$. In a flat
universe containing dark energy and pressureless matter, the
Friedmann equation can be written as $\Omega_{\rm de}+\Omega_m=1$,
where $\Omega_{\rm de}$ and $\Omega_m$ are defined as the ratio of
the densities of dark energy $\rho_{\rm de}$ and matter $\rho_m$ to
the critical density $\rho_{\rm crit}\equiv{3M_{\rm Pl}^2H^2}$,
respectively. From the Friedmann equation, Eqs.~(\ref{eq:rhoq}) and
(\ref{eq:eta}), and the energy conservation equation
$\dot{\rho}_m+3H\rho_m=0$, we can derive a differential equation of
$\Omega_{\rm de}(z)$~\cite{Wei:2007ty}
 \be\label{eq:OqzEoM}
   \frac{d\Omega_{\rm de}}{dz}=-\frac{\Omega_{\rm de}}{1+z}\left(1-\Omega_{\rm de}\right)\left[3-\frac{2\left(1+z\right)}{n}\sqrt{\Omega_{\rm de}}\right],
 \ee
where $z=a^{-1}-1$ is redshift. Furthermore, combining
Eqs.~(\ref{eq:rhoq}) and (\ref{eq:eta}) with the energy conservation
equation $\dot{\rho}_{\rm de}+3H(1+w)\rho_{\rm de}=0$, we can easily
find that the equation-of-state parameter (EOS) of NADE is given
by~\cite{Wei:2007ty}
 \be\label{eq:EOS}
   w=-1+\frac{2}{3n}\frac{\sqrt{\Omega_{\rm de}}}{a}.
 \ee

At the first glance, one might consider that NADE is a two-parameter
model, since besides parameter $n$ the model has another free
parameter $\Omega_{m0}$ coming from the natural initial condition
$\Omega_{\rm de0}=1-\Omega_{m0}$ in solving Eq.~(\ref{eq:OqzEoM})
(note that the subscript ``0" denotes the present value of the
corresponding quantity, hereafter). However, as shown in
Ref.~\cite{Wei:2007xu}, the NADE model is actually a
single-parameter model in practice, thanks to its special analytic
feature $\Omega_{\rm de}=n^2a^2/4$ in the matter-dominated epoch. To
obtain this relation, we can consider the matter-dominated epoch, in
which $H^2\propto\rho_m\propto{a^{-3}}$. From Eqs.~(\ref{eq:rhoq})
and (\ref{eq:eta}), we obtain $\rho_{\rm de}\propto{a^{-1}}$. Then,
from the energy conservation equation $\dot{\rho}_{\rm
de}+3H(1+w)\rho_{\rm de}=0$, we have $w=-2/3$. Comparing $w=-2/3$
with Eq.~(\ref{eq:EOS}), we find that $\Omega_{\rm
de}=n^2a^2/4=n^2(1+z)^{-2}/4$. Note that this relation is also one
of the analytic solutions of
 \be\label{eq:reducedOmegaq}
   \frac{d\Omega_{\rm de}}{dz}=\frac{-\Omega_{\rm de}}{1+z}\left[3-\frac{2(1+z)}{n}\sqrt{\Omega_{\rm de}}\right],
 \ee
which is the reduced form of Eq.~(\ref{eq:OqzEoM}), since
$1-\Omega_{\rm de}\simeq 1$ in the matter-dominated epoch. Thus,
once the value of $n$ is given, Eq.~(\ref{eq:OqzEoM}) can be
numerically solved by using $\Omega_{\rm de}(z_{\rm
ini})=n^2(1+z_{\rm ini})^{-2}/4$ at any $z_{\rm ini}$ deep into the
matter-dominated epoch ($z_{\rm ini} = 2000$ is chosen in
Ref.~\cite{Wei:2007xu}).

By using the initial condition $\Omega_{\rm de}(z_{\rm
ini})=n^2(1+z_{\rm ini})^{-2}/4$ at $z_{\rm ini}=2000$ and the
observational data, Wei and Cai~\cite{Wei:2007xu} firstly
constrained the single-parameter NADE model. After their work, this
initial condition was widely used in the literature; see, e.g.,
Refs.~\cite{Wei:2008rv,Li:2009bn,Li:2010ak}. All results showed that
the only free parameter $n$ could be constrained tightly and the
NADE model could fit the observational data well.

However, the fly in the ointment is that the initial condition
$\Omega_{\rm de}(z_{\rm ini})=n^2(1+z_{\rm ini})^{-2}/4$ at $z_{\rm
ini}=2000$ is obtained by considering a flat universe containing
only dark energy and pressureless matter. So, a natural question we
may ask is whether this initial condition is applicable when we
consider the contribution from radiation. Before we answer this
question, let us first see why we need this discussion.

We all know that the cosmic microwave background (CMB) and
large-scale structure (LSS) observations play an essential role in
testing the cosmological model and constraining its basic
parameters. Generally, we might need to use the full data of CMB
(CMB temperature and polarization power spectra) and LSS (matter
power spectrum) to perform a global fitting. However, such a fitting
consumes a large amount of time and power. As an alternative, two
methods are widely utilized: (i) using the shift parameter $R$ from
CMB~\cite{Bond:1997wr,Wang:2006ts} and distance parameter $A$ of the
BAO measurement~\cite{Eisenstein:2005su}, (ii) employing the
distance prior including $R$, $l_A$ and $z_*$ from CMB~\cite{WMAP7}
and $r_s(z_d)/D_V(z)$ from BAO measurement of Sloan Digital Sky
Survey (SDSS)~\cite{Percival:2009xn}. Here $r_s(z) $ is the comoving
sound horizon whose fitting formula is given by
 \be\label{eq:rs}
  r_s(z)=\frac{1} {\sqrt{3}} \int_0^{1/(1+z)}\frac{da}
  {a^2H(a)\sqrt{1+(3\Omega_{b0}/4\Omega_{\gamma0})a}},
 \ee
where the present photon density parameter
$\Omega_{\gamma0}=2.469\times10^{-5}h^{-2}$ (for $T_{\rm cmb}=2.725$
K) with $h$ the Hubble constant $H_0$ in units of 100 km s$^{-1}$
Mpc$^{-1}$, and $\Omega_{b0}$ is the present baryon density
parameter. For the other quantities mentioned in the method (ii), we
will illustrate them in detail in Sec.~\ref{sec3}. There is no doubt
that the method (ii) encodes more information of the CMB and LSS
data. Note that the distance prior ($R$, $l_A$, $z_*$), as shown in
Ref.~\cite{WMAP7}, is applicable only when the model in question is
based on the standard Friedmann-Robertson-Walker (FRW) universe with
pressureless matter, radiation, dark energy, and spatial curvature.
Since the integral in Eq.~(\ref{eq:rs}) involves the early
radiation-dominated epoch, we need to consider the contribution of
radiation when utilizing $r_s(z_d)/D_V(z)$ of the BAO measurement.
To sum up, in order to use the method (ii), we need to consider the
contribution of radiation besides the pressureless matter and dark
energy.

In the following, we will show that the initial condition
$\Omega_{\rm de}(z_{\rm ini})=n^2(1+z_{\rm ini})^{-2}/4$ at $z_{\rm
ini}=2000$ needs to be amended to accommodate method (ii) in using
the CMB and BAO data. Then, with the revised initial condition and
the current observational data including type Ia supernovae (SNIa),
CMB and BAO, we will constrain the NADE model in a flat universe
with dark energy, matter, and radiation.

\section{New initial condition}\label{sec2}

In a flat universe with dark energy, matter, and radiation, the
Friedmann equation reads
 \be\label{eq:Friedmann3}
  \Omega_{\rm de}+\Omega_m+\Omega_r=1,
 \ee
where $\Omega_r$ is the ratio of the energy density of radiation
$\rho_r$ to the critical density $\rho_{\rm crit}$. Using
Eqs.~(\ref{eq:rhoq}), (\ref{eq:eta}), and (\ref{eq:Friedmann3}), and
combining the energy conservation equations
$\dot{\rho}_m+3H\rho_m=0$ and $\dot{\rho}_r+4H\rho_r=0$, we can
easily derive the differential equation of $\Omega_{\rm de}(z)$,
 \be\label{eq:OqzEoMr}
  \frac{d\Omega_{\rm de}}{dz}=\frac{-\Omega_{\rm de}}{1+z}\left(1-\Omega_{\rm de}\right)\left[3+F(z)-\frac{2(1+z)}{n}\sqrt{\Omega_{\rm de}}\right],
 \ee
where
$F(z)\equiv\frac{\Omega_{r0}(1+z)}{\Omega_{m0}+\Omega_{r0}(1+z)}$.

To solve Eq.~(\ref{eq:OqzEoMr}), of course, we may use the initial
condition $\Omega_{\rm de0}=1-\Omega_{m0}-\Omega_{r0}$. However,
such a treatment will add an extra parameter $\Omega_{m0}$ to the
NADE model, as mentioned above (note that one usually fixes
$\Omega_{r0}=\Omega_{\gamma0}(1 + 0.2271N_{\rm eff})$, and the
standard value 3.04 of the effective number of neutrino species
$N_{\rm eff}$ is required~\cite{WMAP7}). Moreover, it has been shown
in Ref.~\cite{Li:2010ak} that such a two-parameter NADE model cannot
be constrained well by observational data (e.g., $137.702<n<337.974$
at the $1\sigma$ level). On the other hand, if we expect that
$\Omega_{\rm de}(z_{\rm ini})=n^2(1+z_{\rm ini})^{-2}/4$ at $z_{\rm
ini}=2000$ is able to be used as the initial condition in solving
Eq.~(\ref{eq:OqzEoMr}), we must require it at least to satisfy the
equation
 \be\label{eq:reducedOmegaqr}
  \frac{d\Omega_{\rm de}}{dz}=\frac{-\Omega_{\rm de}}{1+z}\left[3+F(z)-\frac{2(1+z)}{n}\sqrt{\Omega_{\rm de}}\right],
 \ee
which is the reduced form of Eq.~(\ref{eq:OqzEoMr}). Here
$1-\Omega_{\rm de}\simeq 1$ in the matter-dominated epoch is used.
Comparing Eq.~(\ref{eq:reducedOmegaqr}) with
Eq.~(\ref{eq:reducedOmegaq}), we can find that the above condition
depends on $F(z)\ll1$ in the matter-dominated epoch, since
$\Omega_{\rm de}=n^2(1+z)^{-2}/4$ satisfies
Eq.~(\ref{eq:reducedOmegaq}) accurately. However, from the
definition
$F(z)\equiv\frac{\Omega_{r0}(1+z)}{\Omega_{m0}+\Omega_{r0}(1+z)}$,
it may have a visible value because of $z\gg 1$ in the
matter-dominated epoch, even though $\Omega_{r0}\ll\Omega_{m0}$. In
fact, we can check that $F(2000)=0.3850$ if we choose
$\Omega_{m0}=0.1334\times{h^{-2}}$ according to the recent WMAP
observations~\cite{WMAP7}. Thus the existence of the non-ignorable
$F(z)$ indicates that we need to find a new initial condition to
solve Eq.~(\ref{eq:OqzEoMr}).

Fortunately, it is not that hard to obtain the new initial condition
applicable for Eq.~(\ref{eq:OqzEoMr}). Let us consider an epoch when
the density of dark energy can be ignored; but we do not need to
distinguish the matter-dominated or the radiation-dominated epoch;
then we have $H^2\propto\Omega_{m0}a^{-3}+\Omega_{r0}a^{-4}$. From
the definition of the conformal age of the universe, we have
$\eta\propto\sqrt{\Omega_{m0}a+\Omega_{r0}}-\sqrt{\Omega_{r0}}.$
Then, from Eq.~(\ref{eq:rhoq}), we can obtain
 \be\label{eq:rhoqvalue}
   \rho_{\rm de}\propto\left(\sqrt{\Omega_{m0}a+\Omega_{r0}}-\sqrt{\Omega_{r0}}\right)^{-2}.
 \ee
Combining Eq.~(\ref{eq:rhoqvalue}) with the energy conservation
equation $\dot{\rho}_{\rm de}+3H(1+w)\rho_{\rm de}=0$, we can obtain
the EOS of dark energy at this epoch,
 \be\label{eq:EoSvalue}
   w=-\frac{2}{3}+\frac{1}{3}\sqrt{\frac{\Omega_{r0}}{\Omega_{m0}a+\Omega_{r0}}}.
 \ee
Then, comparing Eq.~(\ref{eq:EoSvalue}) with Eq.~(\ref{eq:EOS}), we
can easily obtain
  \be\label{eq:Omegaqvaluez}
     \Omega_{\rm de}=\frac{n^2(1+z)^{-2}}{4}\left(1+\sqrt{F(z)}\right)^2.
  \ee
Furthermore, one can also check that Eq.~(\ref{eq:Omegaqvaluez}) is
an analytic solution of Eq.~(\ref{eq:reducedOmegaqr}). As the
evolution of $\Omega_{\rm de}$ satisfies Eq.~(\ref{eq:Omegaqvaluez})
both in the matter-dominated and the radiation-dominated epochs, we
can use Eq.~(\ref{eq:Omegaqvaluez}) at any $z_{\rm ini}$ in these
two epochs as the initial condition to solve Eq.~(\ref{eq:OqzEoMr})
numerically. In our work, we follow Ref.~\cite{Wei:2007xu} and still
choose $z_{\rm ini}=2000$. Then, a new initial condition
$\Omega_{\rm de}(z_{\rm ini})=\frac{n^2(1+z_{\rm
ini})^{-2}}{4}\left(1+\sqrt{F(z_{\rm ini})}\right)^2$ at $z_{\rm
ini}=2000$ is available for the NADE model in a flat universe with
dark energy, pressureless matter, and radiation. In fact, this new
initial condition is also valid in a non-flat universe, since the
spatial curvature $\Omega_k$ is much smaller than $\Omega_m$ or
$\Omega_r$ at $z=2000$. It is interesting to make a comparison
between the new initial condition and the old one $\Omega_{\rm
de}(z_{\rm ini})=n^2(1+z_{\rm ini})^{-2}/4$ at $z_{\rm ini}=2000$.
Obviously, their difference comes from the term
$\left(1+\sqrt{F(z_{\rm ini})}\right)^2$ whose value is 2.626 at
$z_{\rm ini}=2000$, which means that the density of dark energy at
$z=2000$ given by the new initial condition is about 2.6 times
larger than that given by the original initial condition.

Up to now, we have discussed the new initial condition
theoretically. Next, we test the new initial condition by using the
observational data to constrain the NADE model with the new initial
condition. To achieve this, we will use a Markov chain Monte-Carlo
(MCMC) method. However, before doing this, we still need to overcome
a technical difficulty. It is well known that we need to give
initial free parameters to launch the MCMC. For our work, as our
purpose of finding the new initial condition is to keep the
single-parameter feature of the NADE model, we only give an initial
value of the single free parameter $n$. Here, the difficulty is how
to numerically solve Eq.~(\ref{eq:OqzEoMr}) using the new initial
condition $\Omega_{\rm de}(z_{\rm ini})=\frac{n^2(1+z_{\rm
ini})^{-2}}{4}\left(1+\sqrt{F(z_{\rm ini})}\right)^2$ at $z_{\rm
ini}=2000$, as we only know the initial value of $n$ but have no
idea about $\Omega_{m0}$ (the value of $\Omega_{r0}$ is fixed as
mentioned above). Note that both Eq.~(\ref{eq:OqzEoMr}) and the new
initial condition explicitly contain parameter $\Omega_{m0}$.

For this problem, actually we have many methods. Here, we introduce
two methods. The first one is treating the non-independent parameter
$\Omega_{m0}$ as a variable. Thus, Eq.~(\ref{eq:OqzEoMr}) becomes a
partial differential equation,
 \be\label{eq:OqzEoMrPartial}
  \frac{\partial\ln\Omega_{\rm de}}{\partial{z}}
  =\frac{\Omega_{\rm de}-1}{1+z}\left[3+\frac{\Omega_{r0}(1+z)}{\Omega_{m0}+\Omega_{r0}(1+z)}-\frac{2(1+z)}{n}\sqrt{\Omega_{\rm de}}\right],
 \ee
where $\Omega_{\rm de}=\Omega_{\rm de}(z,~\Omega_{m0})$ is a
function of the two variables $z$ and $\Omega_{m0}$. Thus, giving a
value of $n$ and a range of $\Omega_{m0}$, we can numerically solve
the partial differential equation~(\ref{eq:OqzEoMrPartial}) by using
the new initial condition $\Omega_{\rm de}(z_{\rm
ini},~\Omega_{m0})=\frac{n^2(1+z_{\rm
ini})^{-2}}{4}\left(1+\sqrt{F(z_{\rm ini},~\Omega_{m0})}\right)^2$
at $z_{\rm ini}=2000$. Then, with the result of $\Omega_{\rm
de}(z,~\Omega_{m0})$, we can obtain the value of $\Omega_{m0}$ by
numerically finding the root of the equation $\Omega_{\rm
de}(0,~\Omega_{m0})=1-\Omega_{m0}-\Omega_{r0}$. Substituting
$\Omega_{m0}$ back into $\Omega_{\rm de}(z,~\Omega_{m0})$, we can
obtain the evolution of $\Omega_{\rm de}(z)$.

We can also use a numerical iterative method, namely, generating a
sequence $\{\Omega_{m0}^{(i)}\}$ by the iterative formula
$\Omega_{m0}^{(l+1)}=1-\Omega_{\rm
de}(0)|_{\Omega_{m0}^{(l)}}-\Omega_{r0}$. Here, $\Omega_{\rm
de}(z)|_{\Omega_{m0}^{(l)}}$ is the numerical solution of
Eq.~(\ref{eq:OqzEoMr}) with $\Omega_{m0}=\Omega_{m0}^{(l)}$. Thus,
for the current value of $\Omega_{m0}^{(l)}$, a new value
$\Omega_{m0}^{(l+1)}$ can be obtained from the iterative formula. We
can choose the initial value $\Omega_{m0}^{(0)}=0.27$ and set a
termination condition for the iteration, such as
$\left|\Omega_{m0}^{(l+1)}-\Omega_{m0}^{(l)}\right|\leq{\epsilon}$
with $\epsilon$ a small quantity. Our practice shows that its
convergence speed is very fast, and generally, the number of the
iteration is less than 5 for $\epsilon=10^{-6}$.

Using the two methods mentioned above, we find the evolution of
$\Omega_{\rm de}(z)$ for a given value of the single free parameter
$n$. Then, the dimensionless Hubble expansion rate is given by
 \be\label{eq:DimLessH}
   E(z)\equiv {H(z)\over
   H_0}=\left[\frac{\Omega_{m0}(1+z)^3+\Omega_{r0}(1+z)^4}{1-\Omega_{\rm de}(z)}\right]^{1/2}.
 \ee

\section{Observational data and results}\label{sec3}

In this section, we constrain the NADE model with the new initial
condition by using the data from Union2 SNIa (557
data)~\cite{Amanullah:2010vv} and observations of CMB from 7-year
WMAP~\cite{WMAP7} and BAO from SDSS DR7~\cite{Percival:2009xn}.


The data of the 557 Union2 SNIa are compiled in
Ref.~\cite{Amanullah:2010vv}. The theoretical distance modulus is
defined as
 \be\label{eq:mu}
   \mu_{\rm th}(z_i)\equiv5\log_{10}D_L(z_i)+\mu_0,
 \ee
where $\mu_0\equiv42.38-5\log_{10} h$ and the Hubble-free luminosity
distance is
 \be\label{eq:DL}
   D_L(z)=(1+z)\int_0^z \frac{dz'}{E(z',{\bm\theta})},
 \ee
with ${\bm\theta}$ denoting the model parameters. Correspondingly,
the $\chi^2$ function for the 557 Union2 SNIa data is given by
 \be\label{eq:chi2SN}
  \chi^2_{\rm SN}({\bm\theta})=\sum\limits_{i=1}^{557}\frac{\left[\mu_{\rm obs}(z_i)-\mu_{\rm th}(z_i)\right]^2}{\sigma^2(z_i)},
 \ee
where $\sigma$ is the corresponding $1\sigma$ error of distance
modulus for each supernova. The parameter $\mu_0$ is a nuisance
parameter and we can expand Eq.~(\ref{eq:chi2SN}) as
 \be\label{eq:chi2SNTylor}
   \chi^2_{\rm SN}({\bm\theta})=A({\bm\theta})-2\mu_0B({\bm\theta})+\mu_0^2C,
 \ee
where $A({\bm\theta})$, $B({\bm\theta})$ and $C$ are defined in
Ref.~\cite{Nesseris:2005ur}.
Evidently, Eq.~(\ref{eq:chi2SNTylor}) has a minimum for $\mu_0=B/C$
at
 \be\label{eq:minchi2SN}
   \tilde{\chi}^2_{\rm SN}({\bm\theta})=A({\bm\theta})-\frac{B({\bm\theta})^2}{C}.
 \ee
Since $\chi^2_{\rm SN,\,min}=\tilde{\chi}^2_{\rm SN,\,min}$, instead
of minimizing $\chi_{\rm SN}^2$ we will minimize
$\tilde{\chi}^2_{\rm SN}$ which is independent of the nuisance
parameter $\mu_0$.

For the observational data of CMB and BAO, we have mentioned in
Sec.~\ref{sec1} that two simple but efficient methods are often used
instead of using their full data to perform a global fitting. Method
(ii) using distance prior ($R$, $l_A$, $z_*$) from the CMB and
$r_s(z_d)/D_V(z)$ at $z=0.2$ and $0.35$ from the BAO contains more
information of CMB and BAO observations but requires considering the
contribution from the radiation. Since our new initial condition is
mainly designed for this requirement, we will adopt the method (ii)
in our work.

The ``WMAP distance prior'' is given by the 7-year WMAP
observations~\cite{WMAP7}. This includes the ``acoustic scale''
$l_A$, the ``shift parameter'' $R$, and the redshift of the
decoupling epoch of photons $z_*$. The acoustic scale $l_A$
describes the distance ratio $D_A(z_*)/r_s(z_*)$, defined as
 \be\label{eq:lA}
   l_A\equiv (1+z_*){\pi D_A(z_*)\over r_s(z_*)},
 \ee
where a factor of $(1+z_*)$ arises because $D_A(z_*)$ is the proper
angular diameter distance, whereas $r_s(z_*)$ is the comoving sound
horizon at $z_*$ and its fitting formula is given by
Eq.~(\ref{eq:rs}). We fix $\Omega_{b0}=0.02246\times h^{-2}$, which
is given by the 7-year WMAP observations~\cite{WMAP7}. We use the
fitting function of $z_*$ proposed by Hu and
Sugiyama~\cite{Hu:1995en}
 \be\label{eq:zstar}
   z_*=1048[1+0.00124(\Omega_{b0}h^2)^{-0.738}][1+g_1(\Omega_{m0}h^2)^{g_2}],
 \ee
where
 \be\label{eq:91}
   g_1=\frac{0.0783(\Omega_{b0}h^2)^{-0.238}}{1+39.5(\Omega_{b0}h^2)^{0.763}},\quad
   g_2=\frac{0.560}{1+21.1(\Omega_{b0}h^2)^{1.81}}.
 \ee
The shift parameter $R$ is responsible for the distance ratio
$D_A(z_*)/H^{-1}(z_*)$, given by~\cite{Bond97}
 \be\label{eq:R}
   R(z_*)\equiv\sqrt{\Omega_{m0}H_0^2}(1+z_*)D_A(z_*).
 \ee
Following Ref.~\cite{WMAP7}, we use the prescription for using the
WMAP distance prior. Thus, the $\chi^2$ function for the CMB data is
 \be\label{eq:chiCMB}
   \chi_{\rm CMB}^2=(x^{\rm th}_i-x^{\rm obs}_i)(C_{\rm CMB}^{-1})_{ij}(x^{\rm th}_j-x^{\rm obs}_j),
 \ee
where $x_i=(l_A, R, z_*)$ is a vector, and $(C_{\rm CMB}^{-1})_{ij}$
is the inverse covariance matrix. The 7-year WMAP
observations~\cite{WMAP7} give the maximum likelihood values:
$l_A(z_*)=302.09$, $R(z_*)=1.725$, and $z_*=1091.3$. The inverse
covariance matrix is also given in Ref.~\cite{WMAP7}
 \bq\label{eq:incovmat}
   (C_{\rm CMB}^{-1})=\left(\begin{array}{ccc}
   2.305 & 29.698 & -1.333 \\
   29.698& 6825.27 & -113.180 \\
   -1.333& -113.180 &  3.414 \\
   \end{array}\right).
 \eq


We use the BAO data from SDSS DR7~\cite{Percival:2009xn}. The
distance ratio ($d_z$) at $z=0.2$ and $z=0.35$ are
 \be\label{eq:d0.2d0.35}
   d_{0.2}=\frac{r_s(z_d)}{D_V(0.2)},~~
   d_{0.35}=\frac{r_s(z_d)}{D_V(0.35)},
 \ee
where $r_s(z_d)$ is the comoving sound horizon at the baryon drag
epoch~\cite{Eisenstein:1997ik}, and
 \be\label{eq:Dv}
   D_V(z)=\left[\left(\int_0^z\frac{dz'}{H(z')}\right)^2\frac{z}{H(z)}\right]^{1/3}
 \ee
encodes the visual distortion of a spherical object due to the non
Euclidianity of a FRW spacetime. The inverse covariance matrix of
BAO is
 \bq\label{eq:incovmatBAO}
   (C^{-1}_{\rm BAO}) & = & \left(\begin{array}{ccc}
   30124 & -17227 \\
   -17227 & 86977\end{array}\right).
 \eq
The $\chi^2$ function of the BAO data is constructed as
 \be\label{eq:chi2BAO}
   \chi_{\rm BAO}^2=(d_i^{\rm th}-d_i^{\rm obs})(C_{\rm BAO}^{-1})_{ij}(d_j^{\rm th}-d_j^{\rm obs}),
 \ee
where $d_i=(d_{0.2}, d_{0.35})$ is a vector, and the BAO data we use
are $d_{0.2}=0.1905$ and $d_{0.35}=0.1097$.

\begin{figure}[htbp]
\begin{center}
\includegraphics[scale=0.55]{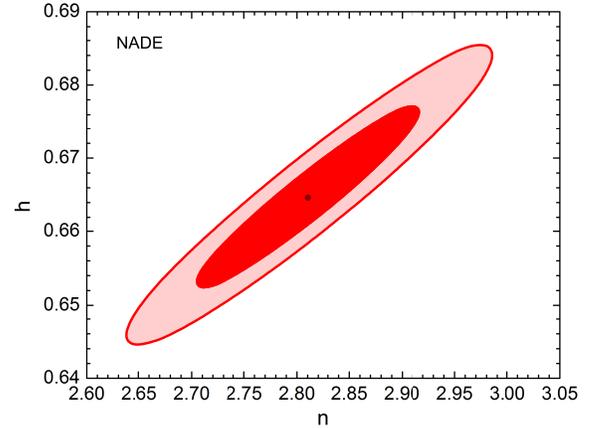}
\caption[]{\small The probability contours at
 $1\sigma$ and $2\sigma$ confidence levels in the $n-h$ plane for the NADE model with the new initial condition.}\label{fig1:NADE}
\end{center}
\end{figure}

The best-fitted parameters are obtained by minimizing the sum
 \be\label{eq:sumchi2}
   \chi^2=\tilde{\chi}^2_{\rm SN}+\chi^2_{\rm CMB}+\chi^2_{\rm BAO}.
 \ee
Using the MCMC method, we finally find the best-fit parameters:
$n=2.810^{+0.113}_{-0.113}$, $h=0.665^{+0.013}_{-0.013}$ at the
$1\sigma$ level, and $n=2.810^{+0.186}_{-0.185}$,
$h=0.665^{+0.022}_{-0.021}$ at the $2\sigma$ level. The best fit
gives $\chi^2_{\rm min}=577.451$ and $\Omega_{m0}=0.268$. In
Fig.~\ref{fig1:NADE}, we plot the contours of $1\sigma$ and
$2\sigma$ confidence levels in the $n$--$h$ plane for the NADE model
with the new initial condition.

The obtained results show that with the new initial condition, the
observational constraints on the single parameter $n$ of NADE model
are fairly tight. Next, we would like to compare the results of this
work with those of previous works. We choose the work in
Ref.~\cite{Li:2010ak} to make a comparison. Note that the work in
Ref.~\cite{Li:2010ak} is also about the observational constraint on
the NADE model but based on the old initial condition, and the
observational data used in Ref.~~\cite{Li:2010ak} also come from the
Union2 SNIa, 7-year WMAP and SDSS DR7, but the method of using CMB
and BAO observations is different from our present work (method (i)
in Ref.~\cite{Li:2010ak} while method (ii) in our present work). So,
the fitting results in these two works are comparable. The
best-fitted $n$ and the corresponding $\Omega_{m0}$ in that work are
$2.886$ and $0.265$, respectively. We see that though the values of
$n$ in the two works are relatively shifted by a small number, the
values of $\Omega_{m0}$ produced by the model are rather similar.
For the $95\%$ limits on $n$, the work of Ref.~\cite{Li:2010ak}
gives $2.723<n<3.055$, and the present work gives $2.625<n<2.996$.
To see the influence of the values of $n$ on the model in the two
works, we will make a comparison with the quantity
$f(z)\equiv\Omega_{\rm de}(z)/\Omega_m(z)$. Using the best-fit
results obtained in the two works, we find that $f(0)_{\rm
new}/f(0)_{\rm old}=0.986$ and $f(2000)_{\rm new}/f(2000)_{\rm
old}=4.362$. So, it is clearly seen that though the difference in
the early epoch is fairly evident, the results produced in the
present time are similar.


\section{Conclusions}\label{sec4}

The NADE model is considered to be a single-parameter model, due to
its special analytical feature $\Omega_{\rm de}=n^2(1+z)^{-2}/4$ in
the matter-dominated epoch. Thus, once the value of the single free
parameter $n$ is given, the differential equation of $\Omega_{\rm
de}(z)$ can be numerically solved by using the initial condition
$\Omega_{\rm de}(z_{\rm ini})=n^2(1+z_{\rm ini})^{-2}/4$ at $z_{\rm
ini}=2000$. However, this initial condition is only applicable in a
flat universe with only dark energy and pressureless matter. That is
to say, when we need to consider the contribution from the
radiation, this initial condition becomes invalid. On the other
hand, some cases indeed need us to consider the contribution of
radiation, for instance, when using the CMB and BAO data to
constrain dark energy models. We mainly have two methods: (i) using
the shift parameter $R$ from the CMB and distance parameter $A$ from
the BAO, (ii) employing the distance prior including $R$, $l_A$ and
$z_*$ from the CMB and $r_s(z_d)/D_V(z)$ from the BAO. Of course,
method (ii) encodes more information of the CMB and the BAO
observations. However, method (ii) requires us to consider the
contribution from the radiation.

Thus, in order to utilize method (ii) to fit CMB and BAO data, we
thoroughly analyzed the NADE model in a flat universe with dark
energy, matter, and radiation. Finally, we found a similar
analytical solution, $\Omega_{\rm
de}=\frac{n^2(1+z)^{-2}}{4}\left(1+\sqrt{F(z)}\right)^2$ with
$F(z)\equiv\frac{\Omega_{r0}(1+z)}{\Omega_{m0}+\Omega_{r0}(1+z)}$,
in the early epoch (matter-dominated or radiation-dominated epoch).
For the initial $z$, we still chose $z_{\rm ini}=2000$. Hence, we
have a new initial condition $\Omega_{\rm de}(z_{\rm
ini})=\frac{n^2(1+z_{\rm ini})^{-2}}{4}\left(1+\sqrt{F(z_{\rm
ini})}\right)^2$ at $z_{\rm ini}=2000$. Furthermore, we found that
this new initial condition is also applicable in a non-flat
universe.

For solving the differential equation of $\Omega_{\rm de}$ before
knowing the value of the non-independent parameter $\Omega_{m0}$, we
provided two methods. The first method is to consider the
non-independent parameter $\Omega_{m0}$ as a variable and use the
new initial condition to solve the partial differential equation of
$\Omega_{\rm de}$. The second method is a numerical iteration
method. Our practice shows that its convergence speed is very fast.
With the two methods, we have constrained the NADE model by using
the new initial condition and the observational data including the
Union2 SNIa, CMB from 7-year WMAP and BAO from SDSS DR7. Our fitting
results show that the values of $n$ and $h$ can be constrained
tightly: $n=2.810^{+0.113}_{-0.113}$, $h=0.665^{+0.013}_{-0.013}$ at
the $1\sigma$ level, and $n=2.810^{+0.186}_{-0.185}$,
$h=0.665^{+0.022}_{-0.021}$ at the $2\sigma$ level. The best fit
gives $\chi^2_{\rm min}=577.451$ and $\Omega_{m0}=0.268$. We believe
that our new initial condition will play a crucial rule in
constraining the NADE model in the future work.


\begin{acknowledgments}
This work was supported by the National Natural Science Foundation
of China (Grant Nos. 10705041, 10975032, 11047112, and 11175042),
the Program for New Century Excellent Talents in University of
Ministry of Education of China (Grant No. NCET-09-0276), and the
National Ministry of Education of China (Grant Nos. N100505001 and
N110405011).
\end{acknowledgments}


\end{document}